\documentclass[11pt,a4paper]{article}
\usepackage{jheppub}
\usepackage{lipsum}
\usepackage{braket}
\usepackage{tensor}
\usepackage{dsfont}
\usepackage{tcolorbox}
\usepackage{caption}
\usepackage{subcaption}
\usepackage{tikz}
\usetikzlibrary{decorations.pathreplacing,calligraphy}
\usetikzlibrary{decorations.pathreplacing}
\usetikzlibrary{decorations.pathmorphing, patterns,shapes}
\usepackage{bm}
\usepackage{caption}
\usepackage{multirow}
\usepackage{subcaption}
\usepackage{ragged2e}
\usepackage{cancel}
\usepackage{comment}
\usepackage{xcolor}
\DeclareCaptionJustification{justified}{\justifying}
\usepackage{hyperref}
\usepackage{amssymb}
\usepackage{tikz}
\usetikzlibrary{decorations.pathreplacing,calligraphy}
\usetikzlibrary{decorations.pathreplacing}
\usetikzlibrary{decorations.pathmorphing, patterns,shapes}
\usepackage{bm}

\newcommand{\op}[1]{\mathcal{#1}}
\newcommand{\de}{\mathrm d}

\title{$\op{PT}$ breaking and RG flows between multicritical Yang-Lee fixed points}

\author[a]{M\'at\'e Lencs\'es}
\affiliation[a]{Wigner Research Centre for Physics, Konkoly-Thege Miklós u. 29-33, H-1121 Budapest, Hungary}
\author[b]{Alessio Miscioscia}
\affiliation[b]{
 Deutsches Elektronen-Synchrotron DESY, Notkestr. 85, 22607 Hamburg, Germany
}%
\author[c]{Giuseppe Mussardo}
 \affiliation[c]{
 SISSA \& INFN, Sezione di Trieste, via Bonomea 265, I-34136, Trieste, Italy
}%
\author[d,e,f]{G\'abor Tak\'acs}
\affiliation[d]{Department of Theoretical Physics, Institute of Physics, Budapest University of Technology and Economics, H-1111 Budapest, M{\H u}egyetem rkp. 3.}
\affiliation[e]{MTA-BME Quantum Correlations Group (ELKH), Institute of Physics, Budapest University of Technology and Economics, H-1111 Budapest, M{\H u}egyetem rkp. 3.}
\affiliation[f]{
BME-MTA Momentum Statistical Field Theory Research Group, Institute of Physics, Budapest University of Technology and Economics, H-1111 Budapest, M{\H u}egyetem rkp. 3.}
\emailAdd{ mate.lencses@gmail.com}
\emailAdd{alessio.miscioscia@desy.de}
\emailAdd{mussardo@sissa.it}
\emailAdd{takacs.gabor@ttk.bme.hu}
\preprint{}
\abstract{
We study a novel class of Renormalization Group flows which connect multicritical versions of the two-dimensional Yang-Lee edge singularity described by the conformal minimal models $\mathcal M(2,2n+3)$. The absence in these models of an order parameter implies that the flows towards and between Yang-Lee edge singularities are all related to the spontaneous breaking of $\op P \op T$ symmetry and comprise a pattern of flows in the space of $\op P \op T$ symmetric theories consistent with the $c$-theorem and the counting of relevant directions. Additionally, we find that while in a part of the phase diagram the domains of unbroken and broken $\op P \op T$ symmetry are separated by critical manifolds of class $\mathcal M(2,2n+3)$, other parts of the boundary between the two domains are not critical. }
\date{\today}

\preprint{DESY-23-035}

\begin{document}
    \maketitle
	\flushbottom
 \section{Introduction and summary}
Renormalization group (RG) flows is a fundamental and fascinating concept in the understanding of scaling and critical behaviour of physical systems with infinitely many degrees of freedom, in particular quantum field theories (QFT) \cite{Wilson:1973jj,Domb:1976bk,Polchinski:1983gv,Amit:1984ms,Binney:1992vn,Cardy:1996xt,Kadanoff:2000xz,2002PhR...363..223B}. The most well-understood universality classes associated with these relativistic quantum theories are the minimal models of two-dimensional conformal field theory \cite{Belavin:1984vu,Cardy:1986ie,Cappelli:1986hf,Zamolodchikov:1986db}. For the unitary cases, the flows originated by these minimal models have been extensively studied, often relying on some  patterns which can be addressed precisely  \cite{Zamolodchikov:1987ti, Ludwig:1987gs,Zamolodchikov:1991vx,Dorey:2000zb,Ahn:2022}. However, much less is known about RG flows which connect non-unitary fixed points corresponding to minimal models of conformal field theories (CFT) where, in the Kac table, there are fields with negative anomalous dimensions and some of the structure constants entering the short distance Operator Product Expansions have imaginary values. A very interesting case of such flows is provided by QFTs associated with critical and multi-critical Yang-Lee edge singularities, i.e. the CFTs behind the density of the accumulated zeros of the grand canonical partition functions of Ising-like models \cite{Yang:1952be,Lee:1952ig}. These models, starting from the seminal work of Michael Fisher \cite{Fisher:1978pf}, have
always attracted a QFT formulation (see, for instance, \cite{Cardy:1985yy,
Cardy:1989fw,Bender:2013qp,Mussardo:2017gao}). There has been lately a renewed interest in their dynamics not just from a purely theoretical point of view \cite{Xu:2022mmw}, but also from the captivating possibility of their experimental realisations, either as a description of decoherence in open systems  \cite{PhysRevLett.109.185701,Matsumoto:2022}, or  via a dynamics of Rydberg atom arrays \cite{Li:2022mjv,Shen:2023tst}. Recently we conjectured \cite{Lencses:2022ira} that conformal minimal models $\mathcal M(2,2n+3)$ generalize the Yang-Lee edge singularity to the multicritical Ising case, providing explicit evidence for the tricritical case previously considered in \cite{vonGehlen:1994rp}. All these RG flows are related to $\op P \op T$ symmetry, which in itself is a topic of considerable interest (c.f. \cite{Bender:98,El-Ganainy:2018,Ashida:2020}, together with the review articles \cite{Bender:2005tb,PhysRevD.70.025001,PhysRevD.98.125003} and references therein). 

\begin{figure}[t]
\centering
\includegraphics[width=0.6\textwidth]{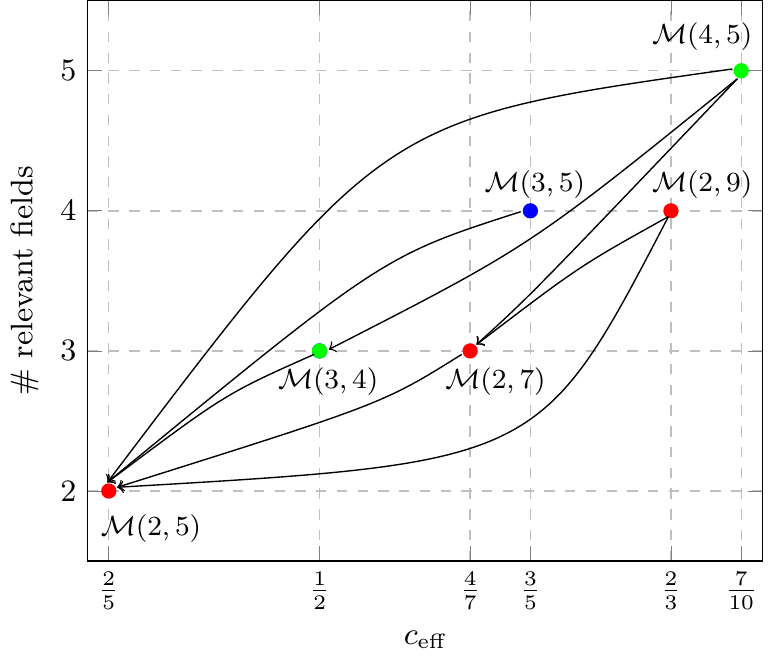}
\caption{Known RG flows between minimal models with effective central charge $c_{\text{eff}}\le 7/10$. Observe that all flows satisfy the $\op P \op T$ symmetric $c-$theorem, and the principle that the number of relevant fields in the ultraviolet is greater than the number of relevant fields in the infrared. Points in red correspond to $\mathcal M(2,2n+3)$ minimal models, the blue dots correspond to the series $\mathcal M(3,q)$ with $q\geq 5$, whereas the green dots represent the unitary minimal models $\mathcal M(p,p+1)$}
\label{fig:RGflows}
\end{figure}

In the case of unbroken $\op P \op T$ symmetry, their spectra are real; the breaking of $\op P \op T$ symmetry is signalled by the appearance of complex conjugate pairs of eigenvalues of the Hamiltonian. As a consequence of $\op P \op T$ symmetry, all these flows satisfy a version of Zamolodchikov's famous $c$-theorem \cite{Zamolodchikov:1986gt}, extended to the $\op P \op T$ symmetric case in \cite{Castro-Alvaredo:2017udm}. Another example of such flows was also considered recently in \cite{Klebanov:2022syt}. 
\newline 

\begin{figure}[t]
\centering
\includegraphics[width=0.6\textwidth]{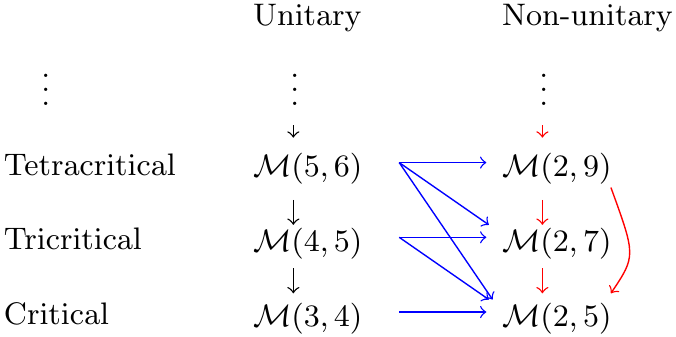}
\caption{The sequence of unitary (Ising) and non-unitary (Yang-Lee) fixed points are shown in columns, with the order of criticality increasing upwards. The black arrows depict integrable flows between multicritical Ising fixed points, the blue arrows are the flows corresponding to Yang-Lee edge singularities \cite{Lencses:2022ira}, while the red arrows are the new flows established in this work.}
\label{fig:RGscheme}
\end{figure}

In this paper we study the scaling region of the minimal models $\mathcal M(2,2n+3)$ spanned by the relevant deformations that leave the Hamiltonian explicitly $\op P \op T$ symmetric, and we present explicit computations for $n=1,2,3$. This study completes the knowledge of RG flows between minimal models with effective central charge $c_\text{eff} \le 7/10$, as depicted in Fig. \ref{fig:RGflows} in the plane spanned by the effective central charge and the number of relevant fields (both these quantities are expected to degrees along RG flows).
\begin{figure}[t]
\centering
\includegraphics[width=0.7\textwidth]{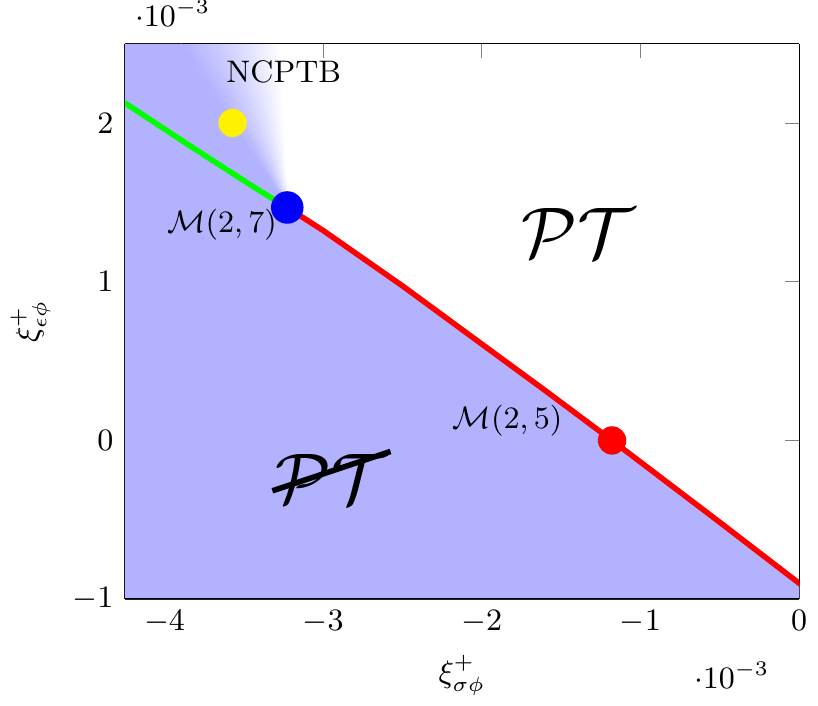}
\caption{Phase diagram of $\op P \op T$ symmetric perturbations of tetracritical Yang-Lee model $\mathcal M(2,9)$. The red line consists of ordinary critical Yang-Lee singularities $\mathcal M(2,5)$, ending in the blue point corresponding to the tricritical Yang-Lee singularity $\mathcal M(2,7)$. The green line corresponds to flows where the first three levels meet simultaneously at some finite value of the volume $R_*$, which diverges when approaching the point $\mathcal M(2,7)$ marked with blue, a behaviour consistent with tricriticality \cite{vonGehlen:1994rp,Lencses:2022ira}. NCTPB means non-critical $\op P \op T$ breaking, and the shading indicates the uncertainty of the corresponding boundary due to lacunae.}
\label{fig:M29phase_diagram}
\end{figure}
We show that the parameter space of the minimal models $\mathcal M(2,2n+3)$ consists of $\op P \op T$ symmetric domains, in which the spectrum is real, and spontaneously broken $\op P \op T$ domains, in which complex conjugate pair of energies appear in the energy spectrum. Our main results are displayed in Figure \ref{fig:M29phase_diagram} which depicts the case relative to the flows starting from  $\mathcal M(2,9)$ ($n=3$): this figure shows that fixed points of the critical Yang-Lee class $\mathcal M(2,5)$ form a sub-manifold in the parameter space, with a boundary corresponding to Yang-Lee tricriticality  $\mathcal M(2,7)$. Generally, in the space of flows starting from $\mathcal M(2,2n+3)$ the fixed points of class $\mathcal M(2,2m+3)$ with $m<n$ are expected to form a sequence of sub-manifolds $M_{m}$ such that the higher multicritical points $M_{m+1}$ form the boundary of $M_{m}$. These multicritical Yang-Lee points connected by these flows themselves arise from RG flows in multicritical Ising models with imaginary magnetic fields \cite{Lencses:2022ira}, as depicted in the scheme in Fig. \ref{fig:RGscheme}.
\newline 

The above scheme of flows is analogous to those between the unitary minimal models $\mathcal{M}(n+1,n+2)$ which were found using perturbation theory for large $n$ \cite{Zamolodchikov:1987ti,Ludwig:1987gs}. However, in the unitary case, the flows between neighbouring fixed points are integrable and so admit an exact solution \cite{1991NuPhB.358..497Z,Zamolodchikov:1991vx}, while the flows between the multicritical Yang-Lee points are not integrable. Additionally, in contrast to the unitary case, these flows are generated by strongly relevant operators and are therefore fully outside the perturbative regime. Consequently, their existence is established here for the first time using non-perturbative numerical methods.  

\section{Multicritical Yang-Lee edge singularities}
The Yang-Lee edge singularity arises in the Ising model with an imaginary magnetic field \cite{Lee:1952ig,Yang:1952be} and its universality class is described by the conformal minimal model $\mathcal M(2,5)$, which captures the critical behaviour of a Landau-Ginzburg theory with potential $V(\varphi) = i \gamma \varphi^3$ \cite{Cardy:1985yy}. It forms the first member of the sequence $\mathcal M(2,2n+3)$ which have $n$ nontrivial relevant scaling fields as shown in Table \ref{TheoryContent} for the cases $n=1,2$ and $3$, where we also indicated the parity of these fields under $\op P \op T$ transformation. Recently we argued \cite{Lencses:2022ira} that the models $\mathcal M(2,2n+3)$ are higher critical versions of the Yang-Lee singularity arising in the multicritical Ising models with imaginary magnetic fields as shown by the blue arrows in Figure \ref{fig:RGscheme}.\newline 

\begin{table}[b!]
\begin{center}
\begin{tabular}{  c || c |c| c| c }
Model & $c_\text{eff}$ &Primary & ($h,\overline h$) & $\op P \op T$  \\ \hline 
\multirow{2}{*}{$\mathcal M(2,5)$} & \multirow{2}{*}{$2/5$}  &
$1$ &(0,0) & even\\
 & &$\varphi$ & (-1/5,-1/5) & odd \\ \hline \hline
\multirow{3}{*}{$\mathcal M(2,7)$ } & \multirow{3}{*}{$4/7$} &
$1$ & (0,0)  & even \\
& &$\psi$ & (-2/7,-2/7) & odd \\
 & &$\rho$ & (-3/7,-3/7) & even \\
 \hline \hline 
 \multirow{4}{*}{$\mathcal M(2,9)$} & \multirow{4}{*}{$2/3$}  &
$1$ & (0,0)  & even \\
& &$\phi$ & (-1/3,-1/3) & odd \\
& &$\sigma$ & (-5/9,-5/9) & even \\
& &$\epsilon$ & (-2/3,-2/3) & odd \\
 \hline \hline 
\end{tabular}
\caption{Effective central charges and primary fields in the minimal models $\mathcal M(2,5)$, $\mathcal M(2,7)$ and $\mathcal M(2,9)$. The parameters ($h,\overline h$) are the right/left conformal dimensions of the fields, while the last column gives their parity under a $\op P \op T$ transformation. }
\label{TheoryContent}
\end{center}
\end{table}

The most general off-critical $\op P \op T$ symmetric deformations of these models are given by the Hamiltonians
\begin{equation}
    \boldsymbol H = \boldsymbol H_{\mathcal M(2,2n+3)}
    +\sum_{\Phi \text{ even}}  \lambda_{\Phi } \int \Phi(x) \ \de x
   +i\sum_{\Phi \text{ odd}}  \lambda_{\Phi } \int \Phi(x) \ \de x \, , 
\label{eq:Hamiltonian}
\end{equation}
where the first term on the RHS is the Hamiltonian of the UV fixed point, $\Phi$ runs over all relevant primary fields, and all coupling constants $\lambda$ are real. Odd and even refer here to the parity of the fields under the $\op P \op T$ symmetry. Note that due to the anti-unitarity of time reversal, all $\op P \op T$-odd perturbing fields must have purely imaginary coefficients. 

The full space of flows generated by $\op P \op T$ symmetric deformations of $\mathcal M(2,2n+3)$ can be parameterised by $n-1$ dimensionless variables. Since all the couplings have nontrivial dimensions, one of them (say $\lambda_\beta$) can be selected to provide an energy/distance scale and then the space of flows can be labelled by dimensionless ratios 
\begin{equation}\label{eq:xidef}
  \xi_{\alpha \beta}^{\pm} = {\lambda_\alpha}/{|\lambda_\beta|^{\frac{h_\alpha-1}{h_\beta-1}}} \ ,   
\end{equation}
 where $\alpha\neq\beta$, and the sign $\pm$ preserves the sign of $\lambda_\beta$ (only matters when the corresponding field is even). The space of flows then splits into domains in which $\op P \op T$ is unbroken and accordingly the spectrum is real, and others in which $\op P \op T$ is spontaneously broken and the spectrum is complex, as illustrated in Fig. \ref{fig:M29phase_diagram} for the case $\mathcal M(2,9)$.
\section{Truncated conformal space}
The RG flows generated by the Hamiltonians in equation \eqref{eq:Hamiltonian} can be studied by putting the system in a finite volume $R$ with $0\leq x\leq R$, and using the volume $R$ as the scale parameter, with $R=0$ corresponding to the short-distance (ultraviolet/UV) fixed point. The presence of a nontrivial long-distance (infrared/IR) fixed point is then signalled by the vanishing of the spectral gap and the following asymptotic behaviour of energy levels as a function of the volume:
\begin{equation}
    E_i(R)-E_0(R)\sim \frac{2\pi\left(\Delta_i-\Delta_0\right)}{R}+\text{subleading corrections}
\label{eq:IR_asymptotics}    
\end{equation}
where $E_0(R)$ is the ground state level, and the different $\Delta_i$ correspond to the sum of conformal weights $h_\Phi+\overline{h}_\Phi$ of scaling operators $\Phi$ of the IR fixed point, with $\Delta_0$ denoting its minimum possible value. The scaling dimensions $\Delta_i$ are the  fingerprint parameters which allow us to uniquely identify the corresponding universality class. 

Here we use the truncated conformal space approach (TCSA) \cite{Yurov:1989yu} to compute the finite volume spectra of the above Hamiltonians, using the chirally factorised algorithm (CFTCSA) introduced in \cite{Horvath:2022zwx}. For all the models $\mathcal M(2,2n+3)$, perturbations with only a single one of the two least relevant fields are integrable \cite{Zamolodchikov:1989hfa,Fateev:1990hy,Cardy:1989fw,Koubek:1991dt}, which for the above explicit cases means $\mathcal M(2,5)$ perturbed by $\varphi$, $\mathcal M(2,7)$ perturbed either by $\psi$ or $\rho$, and  $\mathcal M(2,7)$ perturbed either by $\phi$ or $\sigma$, which allows verifying the accuracy of CFTCSA by comparison to exact predictions from integrability (see next Section). In all these cases, choosing an appropriate sign of the coupling results in a spectrum which has an interpretation in terms of a massive scattering theory guaranteed by $\op P \op T$ symmetry, while the other sign leads to a complex spectrum indicating the spontaneous breaking of $\op P \op T$.

 The particular implementation (CFTCSA) that made use of the chiral factorization of the Hilbert space of a conformal field theory with periodic boundary conditions greatly improves the numerical efficiency of the code \cite{Horvath:2022zwx}. The main source of numerical error is the cutoff used to truncate the Hamiltonian matrix, which can be parameterised by the maximum chiral descendent level retained in the Hilbert space. In our computations, its maximum value was set to  $n_{max} = 10$, which was more than sufficient due to the fact that the perturbations are strongly relevant. We only considered the sector with zero momentum as this was fully sufficient for our purposes.

The only physical parameters are the dimensionless ratios between couplings defined by $\xi_{\alpha\beta}^\pm$ defined in equation \eqref{eq:xidef}, where a selected coupling $\lambda_\beta$ fixes the energy scale to make all quantities dimensionless. Critical points are identified from the volume dependence of the spectrum following \cite{Xu:2022mmw,Lencses:2022ira}. Note that while the numerical value of the critical couplings is (somewhat) sensitive to the cutoff, the qualitative features (such as the different domains of flows and the nature of the lines separating them) are much more robust, as discussed in \cite{Lencses:2022ira}. 
\begin{figure*}[t]
\centering
\begin{subfigure}[t]{.48\textwidth}
  \includegraphics{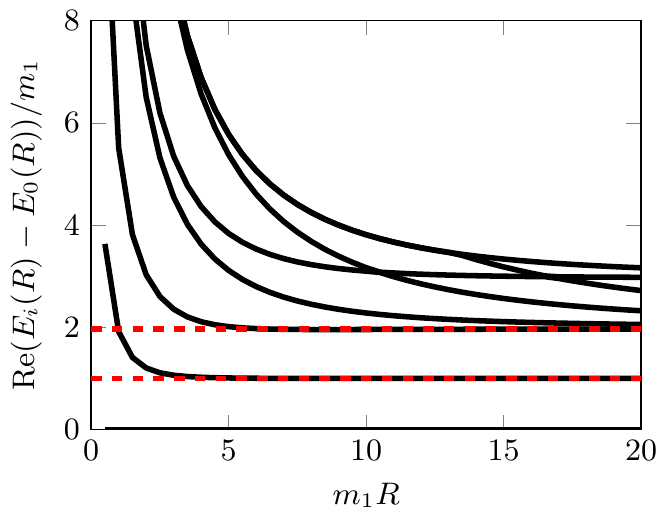}
    \caption{}\label{fig4a}
\end{subfigure}%
\hfill
\begin{subfigure}[t]{.48\textwidth}
  \includegraphics{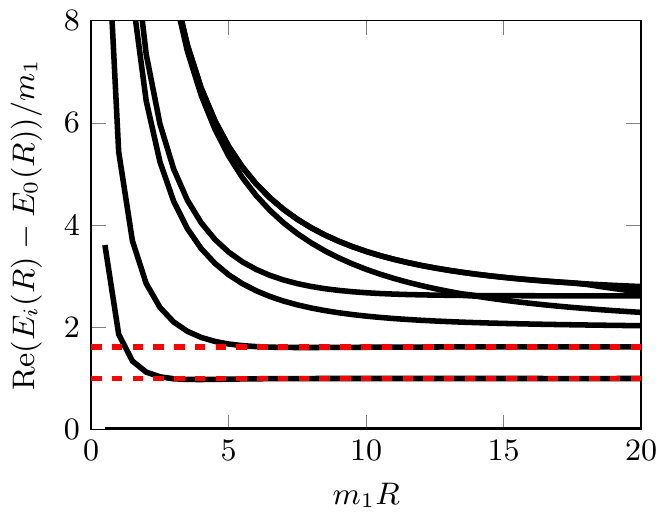}
    \caption{}\label{fig4b}
\end{subfigure}
  \caption{Energy spectra as a function of the volumes (black lines). Left panel (a): the $\Phi_{1,2}$; Right panel (b): the $\Phi_{1,3}$ perturbations of $\mathcal M(2,7)$, compared to the exact predictions for the masses (red dashed lines).}
  \label{Fig_M27_Integrables}
\end{figure*}
\begin{figure*}[t]
\centering
\begin{subfigure}[t]{.48\textwidth}
  \includegraphics{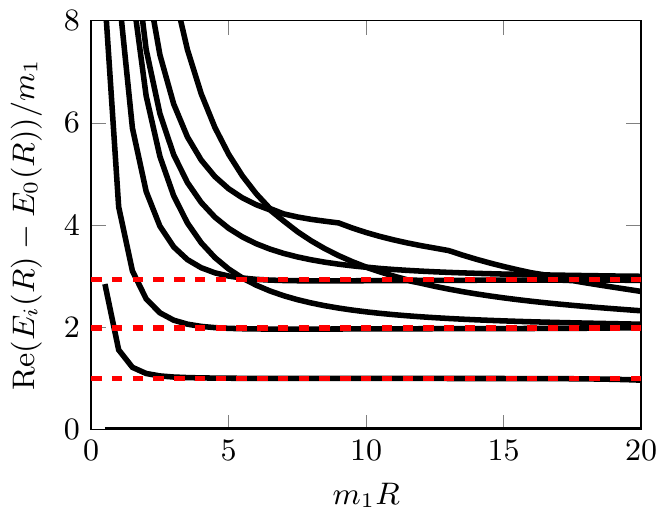}
  \caption{}\label{fig5a}
\end{subfigure}%
\hfill
\begin{subfigure}[t]{.48\textwidth}
  \includegraphics{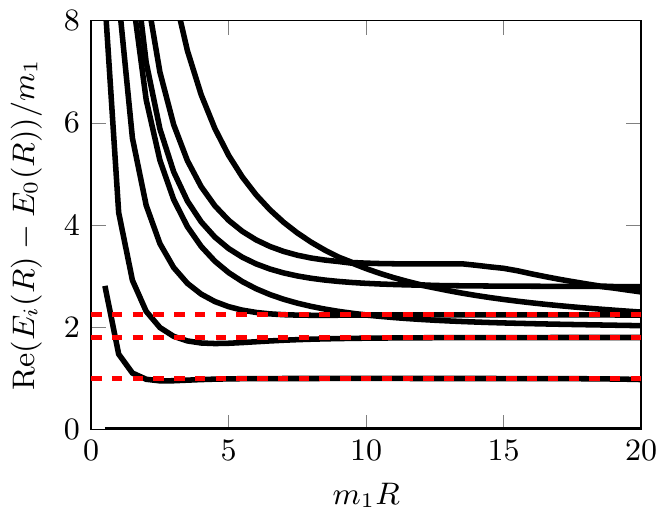}
    \caption{}\label{fig5b}
\end{subfigure}
  \caption{Energy spectra as a function of the volumes (black lines). Left panel (a): the $\Phi_{1,2}$; Right panel (b): the $\Phi_{1,3}$ perturbations of $\mathcal M(2,9)$, compared to the exact predictions for the masses (red dashed lines).}
  \label{Fig_M29_Integrables}
\end{figure*}
\section{Integrable models}

The validity of the CFTCSA Hamiltonian and the accuracy of the spectra can be verified using the existence of integrable directions in the space of perturbations. For the Yang-Lee critical point $\mathcal M(2,5)$, the only relevant perturbation is integrable and well studied in literature \cite{Cardy:1989fw,Yurov:1989yu,Dorey:1996re}. Below we present the results for the finite volume spectra of the integrable deformations of the higher critical Yang-Lee fixed points $\mathcal M(2,7)$ and $\mathcal M(2,9)$ by the relevant fields $\Phi_{1,2}$ and $\Phi_{1,3}$, where the labels refer to their classification according to the Kac table \cite{Belavin:1984vu}. In the conventions of the main text, $\Phi_{1,2}=\varphi$ and $\Phi_{1,3}=\rho$ for $\mathcal M(2,7)$, while $\Phi_{1,2}=\phi$ and $\Phi_{1,3}=\sigma$ for $\mathcal M(2,9)$.

The spectrum of the $\Phi_{1,2}$ perturbation of the minimal model $\mathcal M(2,7)$ is made up by two particles \cite{Koubek:1991dt,Smirnov:1991uw} with masses which are related as 
\begin{equation}
    m_2  = m_1 \frac{\sin \left(\frac{\pi}{9}\right)}{\sin \left(\frac{\pi}{18}\right)} = 1.96962\ldots \ m_1 \,,
\end{equation}
while the coupling of the $\Phi_{1,2}$ field can be expressed with $m_1$ as \cite{Fateev:1993av}
\begin{equation}
    \lambda_\varphi = 0.07855\dots\, m_1^{18/7}\,. 
\end{equation}
Similarly, the spectrum of the $\Phi_{1,3}$ perturbation of $\mathcal M(2,7)$ minimal models contains two particles with the masses related as \cite{1989PhLB..229..243F}
\begin{equation}
    m_2 = m_1 \frac{\sin \left(\frac{2 \pi}{5}\right)}{\sin \left(\frac{\pi}{5}\right)} = 1.61803 \ldots \ m_1\,,
\end{equation}
while the coupling can be expressed in terms of $m_1$ as \cite{Zamolodchikov:1995xk}
\begin{equation}
    \lambda_\rho = 0.04053\dots\, m_1^{20/7}\,, 
\end{equation}
which can again be used to set the units in terms of $m_1$. The spectra extracted from CFTCSA are compared to the exact mass spectrum in Fig. \ref{Fig_M27_Integrables}, which demonstrates the impressive accuracy of the numerical results.

The $\Phi_{1,2}$ and $\Phi_{1,3}$ perturbations of the minimal model $\mathcal M(2,9)$ are completely analogous. The mass spectrum of the $\Phi_{1,2}$ perturbation contains three particles \cite{Koubek:1991dt,Smirnov:1991uw} of masses $m_1$, 
\begin{equation}
    m_2 = m_1 \frac{\sin \left(\frac{\pi}{12}\right)}{\sin \left(\frac{\pi}{24}\right)} = 1.98289 \ldots \ m_1 \,, 
\end{equation}
\begin{equation}
    m_3 = m_1 \frac{\sin \left(\frac{3 \pi}{24}\right)}{\sin \left(\frac{\pi}{24}\right)}  = 2.93185 \dots\, m_1 \ ,
\end{equation}
while the coupling can be expressed in terms of the masses as $\lambda_\phi =  0.08166\dots\, m_1^{8/3}$ \cite{Fateev:1993av}. Similarly, the mass spectrum of the $\Phi_{1,3}$ perturbation contains three particles \cite{1989PhLB..229..243F} with masses $m_1$, 
\begin{equation}
    m_2 = m_1 \frac{\sin \left(\frac{2 \pi}{7}\right)}{\sin \left(\frac{\pi}{7}\right)} = 1.80194\ldots \ m_1  \,,
\end{equation}
\begin{equation}
    m_3 = m_1 \frac{\sin \left(\frac{3 \pi}{7}\right)}{\sin \left(\frac{ \pi}{7}\right)} = 2.24698\ldots \ m_1 \,  , 
\end{equation}
and the mass scale is related to the coupling as $\lambda_\sigma = 0.00989\dots m_1^{28/9}$ \cite{Zamolodchikov:1995xk}. The spectra extracted from CFTCSA are again in excellent agreement with the exact mass spectrum as shown in Fig. \ref{Fig_M29_Integrables}.

\section{The critical $\op P \op T$ boundary}
In the case when the boundary between $\op P \op T$ symmetric and symmetry broken phases is critical, the boundary points correspond to $\op P \op T$ symmetric massless flows. These flows terminate at some IR fixed point satisfying the $\op P \op T$ $c$-theorem \cite{Castro-Alvaredo:2017udm} which states that the effective central charge \footnote{The effective central charge of a CFT is defined by $c_\text{eff} = c-24 h_\text{min}$, where $h_\text{min}$ is the lowest conformal weight in the theory.} of the ultraviolet theory is greater than the effective central charge of the infrared theory, i.e. $c_\text{eff}^{UV}> c_\text{eff}^{IR}$. 
 For a model $\mathcal M(p,q)$, $c_\text{eff} = 1-6/(p q)$ and therefore the above condition implies 
$(p q)_{UV}> (p q)_{IR}$. 

\begin{figure}
\centering
\includegraphics[width=0.7\textwidth]{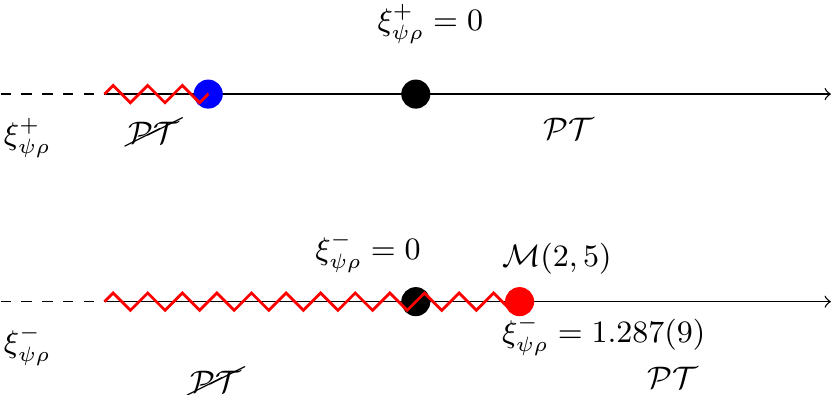}
\caption{Phase diagram for the scaling region of the minimal model $\mathcal M(2,7)$. The black dots correspond to $\lambda_\psi=0$, while the blue dot indicates the non-critical $\op P \op T $ boundary point. The two lines can eventually be joined into a circle by connecting them at $\pm\infty$. The relevant spectra are completely analogous to those shown in Figs. \ref{fig:spectra} and \ref{fig:noncritPTspectrum} for the $\mathcal{M}(2,9)$ case.}
\label{fig:m27phasediagram}
\end{figure}

\begin{figure}[h!]
\centering
  \includegraphics[scale = 1.3]{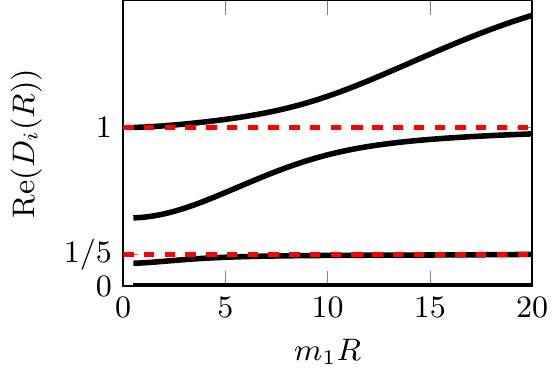}
  \caption{Critical behaviour of the $\mathcal M(2,5)$ universality class in the scaling region of $\mathcal M(2,7)$, (at $ \xi_{\psi \rho}^{-} = 1.2879$) illustrated by plotting the scaling functions $D_i(R)$, with the red dashed lines showing their large volume limits expected from the scaling dimensions in $\mathcal M(2,5)$.}
\label{fig:M27massFlow}
\end{figure}

The simplest $\op P \op T$ invariant RG flow is the usual Yang-Lee singularity in the Ising model flows from $\mathcal M(3,4)$ to $\mathcal M(2,5)$ (for a detailed analysis c.f. \cite{Xu:2022mmw}). For the multi-critical Yang-Lee models considered here, a similar situation occurs in the one-dimensional space of RG flows in $\mathcal M(2,7)$ shown in Fig. \ref{fig:m27phasediagram}. This can be established by scanning the spectrum as a function of the parameter space numerically using CFTCSA, and analysing the spectrum at points corresponding to massless flows in terms of the scaling functions $D_i(R)={R}\left(E_i(R)-E_0(R)\right)/{4\pi}$ which approaches $h_i-h_{min}$ according to Eq. \eqref{eq:IR_asymptotics} in the large $R$ limit \cite{Fonseca:2001dc,Xu:2022mmw,Lencses:2022ira}. The resulting asymptotics can be compared to predictions from the field content of the different universality classes (c.f. Table \ref{TheoryContent}) to identify the IR endpoint of the flow. The only massless flows in the scaling region of the minimal model $\mathcal M(2,7)$ ends in the universality class of the Yang-Lee fixed point as shown in Fig. \ref{fig:M27massFlow}.

\hfill

\begin{figure*}
    \begin{subfigure}[t]{.48\textwidth}
  \includegraphics{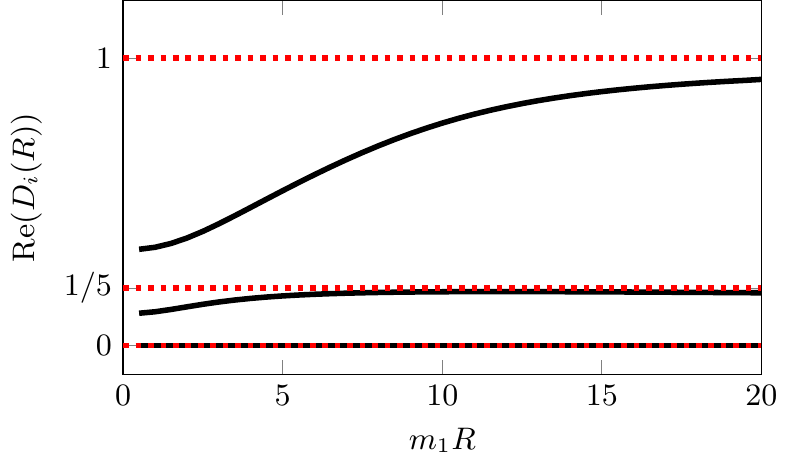}
  \caption{}\label{fig.7a}
\end{subfigure}%
\hfill
\begin{subfigure}[t]{.48\textwidth}
  \includegraphics{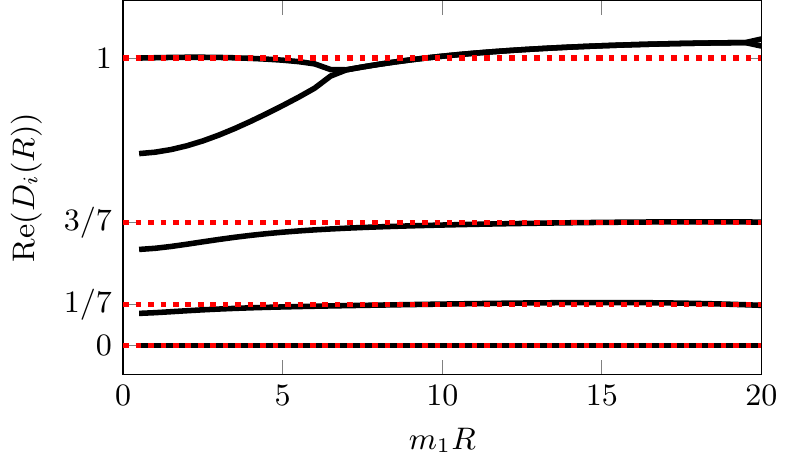}
  \caption{}\label{fig.7b}
\end{subfigure}
    \caption{Representative examples of spectra corresponding to flows classified in Fig. \ref{fig:M29phase_diagram}. Left panel (a): scaling functions $D_i(R)$ $i=1,2,3$ at the point $(\xi_{\sigma \phi}^+, \xi_{\epsilon \phi}^+) = (-0.00118,0)$ along the critical line (red dot in figure \ref{fig:M29phase_diagram}), with horizontal red lines depicting the expected asymptotics for a fixed point of class $\mathcal M(2,5)$. Right panel (b): Scaling functions $D_i(R)$ $i=1,2,3,4,5$ at the endpoint $(\xi_{\sigma \phi}^+, \xi_{\epsilon \phi}^+) = (-0.00322,0.00146)$ (blue dot in figure \ref{fig:M29phase_diagram}),  with horizontal red lines depicting the expected asymptotics for a fixed point of class $\mathcal M(2,7)$. The graphs only show the real part of $D_i(R)$, with imaginary parts occurring only for the lacuna present in the top two levels of the figure in the right panel. Units are in terms of $m_1$ defined as $\lambda_\phi =  0.08166\dots\, m_1^{8/3}$.}
    \label{fig:spectra}
\end{figure*}

The next model $\mathcal M(2,9)$ has a two-dimensional space of flows, and mapping their parameter space using the above method results in Fig. \ref{fig:M29phase_diagram} \footnote{Note that since the field $\phi$ used for normalisation is odd, the diagram does not depend on the sign of the coupling $\lambda_\phi$.}. Representative examples for finite volume spectra determined using CFTCSA results are provided in Fig. \ref{fig:spectra}.

We note that the $c$-theorem alone would allow for the flows to end in the fixed point $\mathcal M(3,5)$ besides $\mathcal M(2,5)$, $\mathcal M(2,7)$; however, this case never occurs, which corresponds to the expectation that the IR fixed point of the RG flow should have less relevant directions than the UV one. In this particular case, the UV fixed point $\mathcal M(2,9)$ has $3$ non-trivial relevant operators which is the same number as in $\mathcal M(3,5)$, while $\mathcal M(2,5)$ and $\mathcal M(2,7)$ have $1$ and $2$, respectively.

\section{Non-critical $\op P \op T$ breaking transition}
\begin{figure*}[t]
\centering
\begin{subfigure}[t]{.45\textwidth}
  \includegraphics[width=0.98\textwidth]{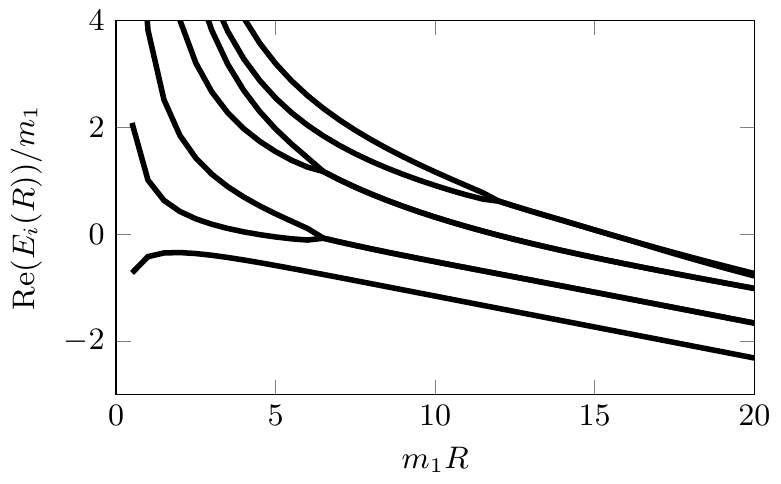}
  \caption{}
\label{fig:noncritPTspectrum}
\end{subfigure}%
\hfill
\begin{subfigure}[t]{.45\textwidth}
  \includegraphics[width=0.98\textwidth]{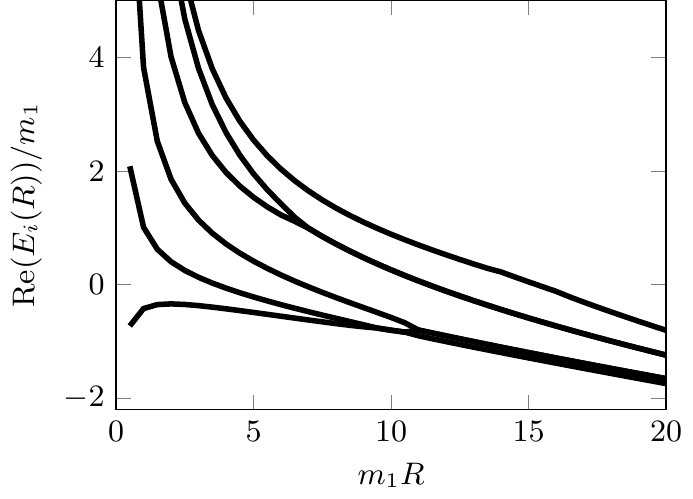}
   \caption{}
    \label{fig:TPM}
\end{subfigure}
  \caption{On the left panel (a): real part of the energy levels $E_i(R)$ for an RG flow corresponding to a non-critical breaking of $\op P \op T$ symmetry, with parameters $(\xi_{\sigma \phi}^+, \xi_{\epsilon \phi}^+) = (-0.00357,0.00201)$ (yellow dot in Fig. \ref{fig:M29phase_diagram}). Energy levels for volumes beyond the branch points form complex conjugate pairs, with non-vanishing imaginary parts. Volume units are  as in Fig \ref{fig:spectra}. On the right panel (b): An example of energy spectra at the line of triple-level meeting in the scaling region of the minimal model $\mathcal M(2,9)$: $(\xi_{\sigma \phi}^+,\xi_{\epsilon \phi}^+) = (-0.00345, 0.001604)$.}
\end{figure*}

We have  also demonstrated the existence of non-critical $\op P \op T$ symmetry-breaking transitions. Note that $\op P \op T$ symmetry breaking has no order parameter; it corresponds to the appearance of complex conjugate energy levels. Approaching from the $\op P \op T$ symmetric side this proceeds via the disappearance of a gap between some pair of adjacent energy levels. The transition is only critical when the levels involved are the ground state and the first excited state, otherwise, it is more reminiscent of the so-called \textit{excited-state quantum phase transitions} (see \cite{Cejnar_2021} for a recent review). 
Similarly to the unitary case, critical lines terminate in a point corresponding to a different universality class; in the examples above, a critical line of Yang-Lee class $\mathcal M(2,5)$ terminates in a point ruled by the tricritical Yang-Lee class $\mathcal M(2,7)$. Beyond the end point the transition becomes non-critical in the sense above. However, identifying a sharp boundary is extremely difficult due to the appearance of so-called \textit{lacunae}, which are pairs of energy levels becoming complex for some finite volume interval \cite{Yurov:1989yu,1997NuPhB.489..557K}. When moving towards the $\op P \op T$ breaking regime, these lacunae grow longer and Fig. \ref{fig:noncritPTspectrum} shows a case where two of them either extend to infinite volume or at least to volumes higher than CFTCSA can handle. It is difficult also to exclude from CFTCSA that some higher excited levels have already gone complex and so in the absence of critical behaviour the precise location of the boundary between the two regimes cannot be reliably determined. 
\newline 

We further expect the existence of a line of triple point meeting in the $\op P \op T$ broken region of the scaling region of the minimal model $\mathcal M(2,7)$. This line eventually ends in the tricritical version of the Yang-Lee minimal fixed point, conjectured to be in the universality class of the minimal model $\mathcal M(2,7)$ \cite{Lencses:2022ira}. It is possible to check, with the TCSA, that this line indeed exists (green line in Fig. \ref{fig:M29phase_diagram}). An example of a (non-critical) triple point meeting is given in Fig. \ref{fig:TPM}.

\section{Discussion} 
The explicit demonstration of RG flows from fixed points $\mathcal M(2,7)$ and $\mathcal M(2,9)$ to those of lower criticality naturally leads to conjecture the existence of flows from $\mathcal M(2,2n+3)$ to all models $\mathcal M(2,2n'+3)$. As shown in Fig. \ref{fig:RGscheme}, these results are fully consistent with the recent conjecture that the universality classes $\mathcal M(2,2n+3)$ are multicritical versions of the Yang-Lee edge singularity \cite{Lencses:2022ira}. Indeed, from the conjecture of \cite{Lencses:2022ira} the existence of the flows studied here is expected because all the fixed points $\mathcal M(2,2n'+3)$  with $n'< n$ lie in the space of $\op P \op T$ symmetric deformations of the unitary multicritical models $\mathcal M(n+1,n+2)$. All these flows satisfy the $\op P \op T$ symmetric extension of the $c$-theorem \cite{Castro-Alvaredo:2017udm} and also satisfy the principle that the IR fixed point of the RG flow must have a smaller number of independent relevant directions than the UV fixed point. An interesting future direction is to attempt the full classification of $\op P \op T$ symmetric RG flows between minimal models, guided by the $\op P \op T$ symmetric $c$-theorem, global symmetries, and the constraint on the number of relevant operators, extending therefore the results summarized in Fig. \ref{fig:RGflows} for the case $c> 7/10$. Another interesting issue is to understand the nature of non-critical $\op P \op T$ breaking, possibly in the framework of excited state phase transitions \cite{Cejnar_2021}. Finally, recent advances \cite{Peng:2015,Li:2022mjv,Matsumoto:2022,Nakayama:2021zcr} indicate the potential experimental relevance of the Yang-Lee phenomenon studied in the present work.

\begin{acknowledgments}
 GM acknowledges the grants Prin 2017-FISI and PRO3 Quantum Pathfinder. The work of ML was supported by the National Research Development and Innovation Office of Hungary under the postdoctoral grant PD-19 No. 132118 and partially by the OTKA Grant K 134946. GT was partially supported by the Ministry of Culture and Innovation and the National Research, Development and Innovation Office (NKFIH) through the OTKA Grant K 138606, and also by the Ministry of Culture and Innovation and the National Research, Development and Innovation Office under Grant Nr. TKP2021-NVA-02. AM is partially supported from the German Research Foundation DFG under Germany’s Excellence Strategy – EXC 2121 Quantum Universe – 390833306 and from the European Union’s Horizon 2020 research. ML, GM and GT are also grateful for the hospitality of KITP Santa Barbara during the  program "Integrability in String, Field, and Condensed Matter Theory", where a part of this work was completed. This collaboration was supported in part by the National Science Foundation under Grant No. NSF PHY-1748958, and by the CNR/MTA Italy-Hungary 2023-2025 Joint Project “Effects of strong correlations in interacting many-body systems and  quantum circuits”.  AM is grateful to the Galileo Galilei Institute for Theoretical Physics in Firenze for hosting him during LACES 2022.
\end{acknowledgments}
	 \bibliographystyle{JHEP}
	 \bibliography{PTB}
\end{document}